\documentclass[a4paper, amsfonts, amssymb, amsmath, reprint, showkeys, nofootinbib, twoside]{revtex4-2}
\usepackage[english]{babel}
\usepackage[utf8]{inputenc}
\usepackage[colorinlistoftodos, color=green!40, prependcaption]{todonotes}
\usepackage[pdftex, pdftitle={Article}, pdfauthor={Author}]{hyperref} 
\usepackage{empheq,etoolbox}
\patchcmd{\subequations}
  {\theparentequation\alph{equation}}
  {\theparentequation.\alph{equation}}
  {}{}
\begin{document}
\title{Signal selective amplification for below-the-threshold stimulus in Fitzhugh-Nagumo neuronal model}

\author{Mariia Sorokina}
    \email[Correspondence email address: ]{k4939b@gmail.com}
    \affiliation{90 Navigation street, Birmingham, UK, B5 4AA}

\date{\today} 

\begin{abstract}
Brain operates at remarkably low signal power. It has been noted that noise may play a constructive role in neural networks and facilitate the  subthreshold signaling. The process of spiking pattern excitation at the characteristic neuronal spiking frequency from the random noisy stimulus remains unexplained. Furthermore, recent research indicates that neuronal processing enables signal selective amplification. Also, there is a growing number of studies indicating the role of coloured noise for spike generation, while noise shaping effect have been recently observed in neuronal networks as well as in single neurons. 
We demonstrate for the first time that noise shaping enables efficient signal amplification and spike excitation for below-the-threshold stimulus. On the example of the seminal FitzHugh-Nagumo (FHN) model we reveal that the complex interplay between the fast and slow FHN variables leads to noise accumulation at the FHN characteristic spiking frequencies, while tails follow a power-law frequency scaling.  We demonstrate that the discovered effect allows selective amplification of signal waveforms  presenting a novel type of amplification process.
\end{abstract}


\maketitle

\section*{Introduction}
The constructive impact of noise on neuronal systems have been extensively studied via the prism of stochastic resonance (SR). Furthermore, SR has been used to explain subthreshold crossings and signal amplification in neuronal models.  However, since mid 1990s it has been reported that the optimum noise intensity in such cases has more complex and intricate dependency on the stimulus frequency than could be explained by the conventional models \cite{SR}. So that, by the end of 1990s there was a surge of interest  to  the role of noise  \cite{NA} and its impact on frequency selectivity in neuronal processing \cite{review}. Most recently the ability of inner ear's cochlea to selectively amplify signals near the characteristic frequency distinguishing them from the noise at the same frequency was revealed \cite{STN}. 

On the other hand there is a growing body of research indicating the impact of coloured noise \cite{CN2}. Recently, noise shaping (NS) effect has been observed in neuronal networks \cite{eckhorn}. Also NS in a single neuron based on the standard FitzHugh-Nagumo (FHN) model was reported \cite{HFT}. This is also the first time  NS is observed in a nonlinear system without a feedback loop. 
While the resonance in two time scales (one deterministic - periodic excitation and another stochastic - noise) is the essence of SR,  NS represents an interplay of two time scales of an input  and a feedback signals. As a result, in SR the noise enhances the switching induced by the periodic input, while in NS the noise is reshaped in frequency, i.e. moved away from the frequencies occupied by the signal.  Thus, in NS the feedback loop essentially acts as a filter and the output noise becomes coloured, due to this by the appropriate  receiver design one can ensure significant improvement in signal reception quality.  

Here on the example of the FHN model we show that noise shaping plays an important role in sub-threshold signal amplification. We explain  a two time-scales interplay of the FHN variables (fast  - voltage and slow - recovery playing a role of a feedback). We show that the spectra of the fast and slow variables   become coloured having $f^{-2}$ and  $f^{-4}$ tails, moreover the spectra exhibit peaks around the frequencies characteristic to the FHN neuronal spiking frequencies (i.e. when excited by a constant stimuli above-the-threshold), which enables selectivity in amplification: though originally at the same frequency the signal is selectively amplified, while the noise is suppressed. Moreover, we demonstrate the waveform selectivity: signals of the characteristic waveform are amplified more. This is a new type of amplification.

\begin{figure*}[!ht]
\centering
\includegraphics[width=\linewidth]{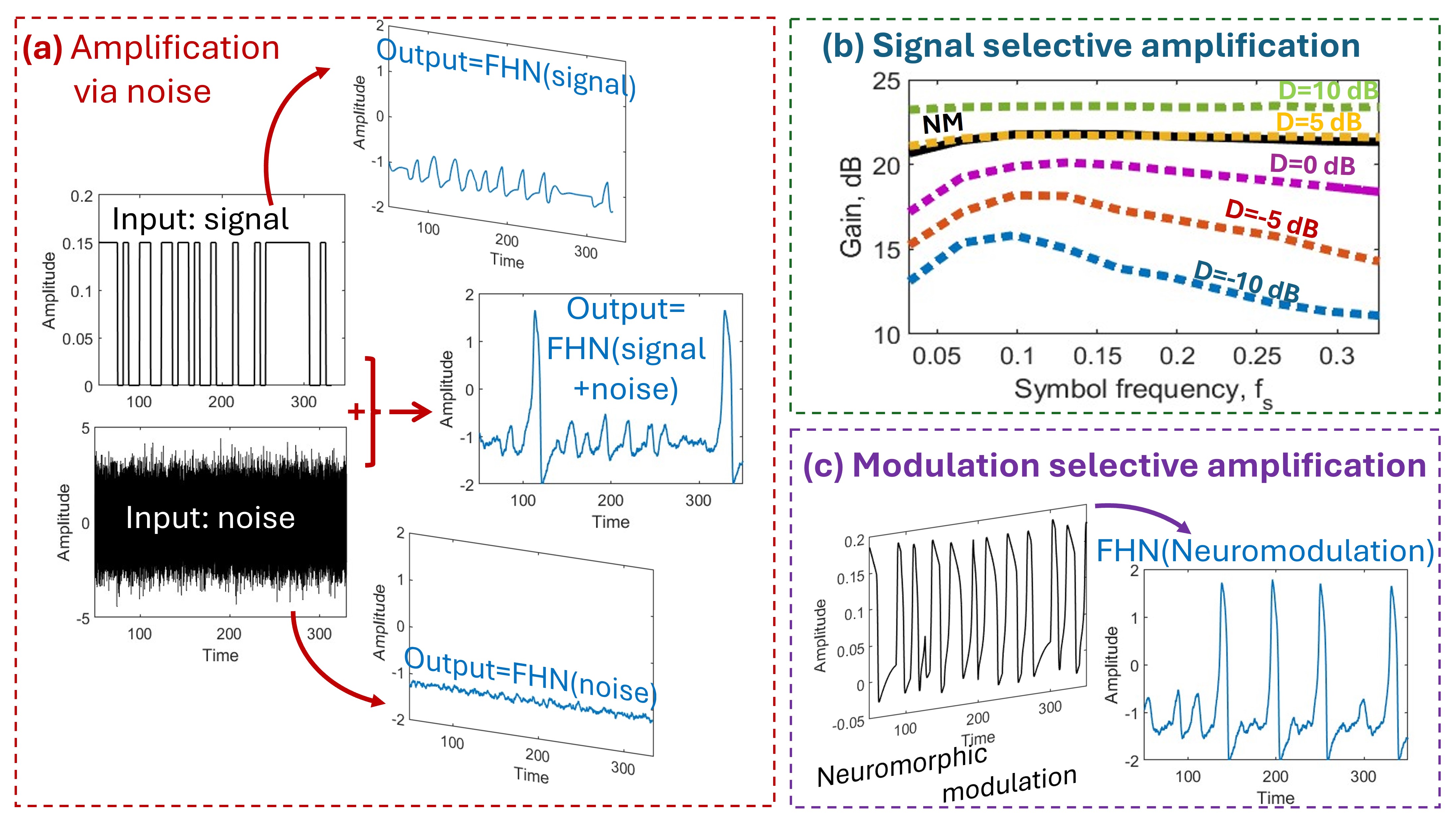}
\caption{\textbf{Signal selective amplification.}  
\textbf{a} The FHN neuron does not generate spiking when excited by either: small (below-the-threshold) signal or noise, yet spiking occurs when \textit{noise is added to the  signal}. The input signal is randomly encoded with On-Off Keying rectangular  modulation having amplitude $A_s=0.15$ and symbol frequency $f_s=0.15$. The noise is AWGN with variance: $D=1$.  \textbf{b} The gain (the ratio of output and input signal powers) is plotted for the aforementioned signal parameters for varied $D$ (coloured dashed lines). The results are: (i) the FHN neuron selectively amplifiers the signal by transferring noise into the signal: more noise - more gain; (ii) the FHN neuron selectively amplifiers signals with neuromorphic modulation (NM): the gain (shown in black solid lines) for NM signal without noise is comparable to the rectangular signal requiring 5 dB noise. \textbf{c} If \textit{neuromorphic modulation} is used the signal excites spiking even for below-the-threshold powers. }
\label{F1}
\end{figure*}

\begin{figure*}[!ht]
\centering
\includegraphics[width=\linewidth]{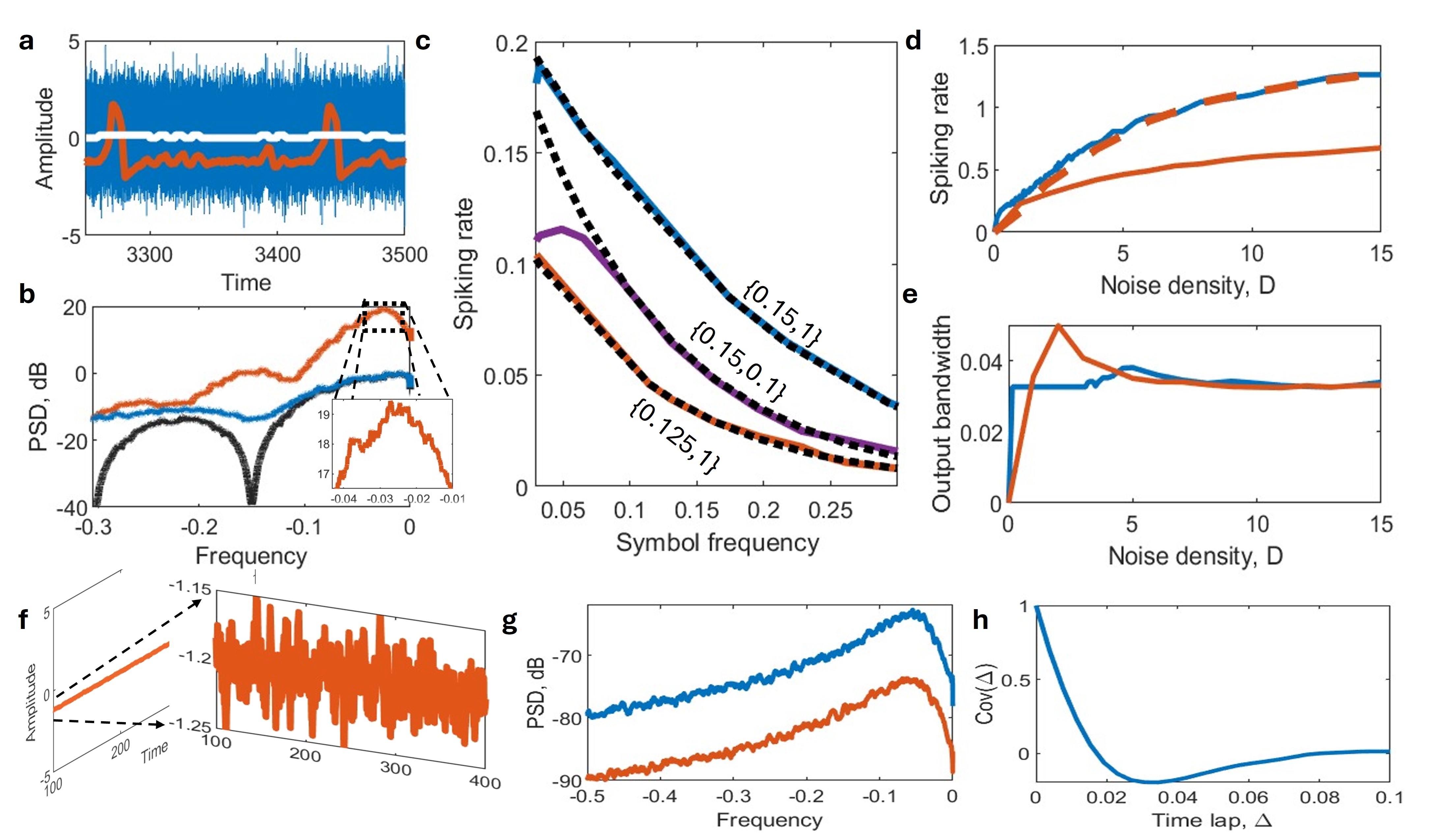}
\caption{\textbf{Amplification via noise shaping.}  
\textbf{a} Spiking (red lines) can be achieved for below-the-threshold signals (white) if noise distortion is large (blue). The input (white) is On-Off Keying rectangular signal with amplitude $A_s=0.15$ and symbol frequency $f_s=0.15$. Noise variance: $D=1$.  \textbf{b} The corresponding power spectral density (PSD) is plotted for output (red) and input (blue) signals, the pure input signal is plotted for reference (black) (for illustration $RBW=5 \cdot 10^{-5}$ is used). The output spectrum is reshaped around the neuronal characteristic frequency $f_0$ as highlighted by the inset. \textbf{c} The corresponding spiking rate is plotted as a function  of symbol frequency  for various input parameters $\{A_s,D\}$ in solid lines. The corresponding asymptotic expressions are shown in dashed lines and demonstrate the exponential dependence on $f_s$. Spiking rate dependency (Fig. 1c) on signal amplitude enables to receive more frequent spiking for "ones" than for "zero"-bits, thus enabling decoding, while the frequency dependency allows   to optimize the input for better efficiency.  The inset highlights the critical point $f_0$, below which  multiple  spikes per symbol occur.  \textbf{d} The spiking rate  for large $D$  saturates to a constant value defined by $f_0/f_s$ - the number of spikes per symbol (see $f_0/f_s=1, 2$ in blue and red solid lines). When normalized by a number of spikes per symbol (see dashed red line) both cases coincide. \textbf{e} The output bandwidth for the cases considered in panel d  demonstrate the preservation of bandwidth at $f_0$ - neuron characteristic frequency value.  \textbf{f} Noise shaping: compare the output in the absence of signal (noise only) (left) and magnified (right) plotted for the same parameters as in panel a, see PSD in \textbf{g}  for $D=1, 0.1$ in blue and red lines. \textbf{h} The normalized covariance of output noise for the varied time lap. Thus, optimizing modulation: symbol frequency for a given amplitude and noise may result in efficient signal amplification for below-the-threshold signals.  }
\label{F1}
\end{figure*}

\section*{Results: Signal selective amplification}
Let us consider the Fitzhugh-Nagumo (FHN) model \cite{FHN1,FHN2}: 
\begin{subequations}
  \begin{empheq}[left=\empheqlbrace]{align}
\dot{v} &=v-\frac{v^3}{3}-w+I \\
\tau \dot{w} &=v-bw+a
  \end{empheq}
  \label{FNH_o}
\end{subequations}
for the output (voltage) $v$ and recovery variable $w$ with fixed model parameters $a=0.8, b=0.7, \ \tau=10$ (unless otherwise specified).

Here we consider the case when an excitation $I$ is below the amplitude threshold. In such case the spikes are not generated (see Fig. 1a), similarly to the case when only noise is fed into the FHN. Though when inserted on their own the small signal and large noise do not excite spiking, when added together - spiking occurs. Thus, the FHN processing results in signal amplification and occurs only in the presence of strong noise (noise is significantly larger then signal). The corresponding gain (the ratio of input (clean) signal to the FHN generated output is plotted in Fig. 1b for various values of $D$ and fixed signal parameters. One can also straightforwardly obtain the gain as a ratio of noisy input power to the output using the values of $D$. Here we chose the aforementioned definition as noise is added intentionally as a part of amplification process.  Here an input signal is a rectangular signal with the amplitude $A_s=0.15$ randomly encoded as "zeros" and "ones"  (On-Off Keying)  with symbol frequency $f_s=0.15$ in Fig.1a and varied in panel Fig. 1b. It is important to note that the effect occurs only for $f_s\geq f_0$, where $f_0$ is the FHN eigen-frequency - characteristic spiking frequency defined as the maximum spiking frequency of FHN when excited by a constant stimuli, see \cite{HFT}. The maximum gain occurs for $f_s=2f_0$ (for the given parameters of FHN: $f_0=0.032$, generally $f_0\simeq 1/(\pi\tau)$ for $\tau\ll 1$, see \cite{HFT}). At higher noise powers the dependence on input symbol frequency becomes less pronounced and the gain growth starts to saturate (compare the gain increase when noise rises from -10 to 0 dB and from 0 to 10 dB levels).  Thus, unlike, conventional amplifiers, where signal and noise are both equally amplified, the FHN processor can selectively amplify a signal by transferring noise power into the the signal. Thus, one can achieve the FHN spiking for below-the-threshold signal powers adding more noise into the input: more noise - more gain.

As we have seen the FHN processor can effectively select signal from noise and use the noise advantageously. Let us now consider the impact of the signal waveform. Next we introduce a concept of neuromorphic waveform. In \cite{HFT} we have noted that the FHN can be considered as a modulator: a nonlinear combination of amplitude shift keying (ASK) and frequency shift keying (FSK). While a combination of those -  AFSK - is an area of active research,  here we have a case of a nonlinear combination with a power threshold with unique advantages in bandwidth (see \cite{HFT}). To receive the FHN waveform, let us feed into the FHN a rectangular waveform with within-the-threshold amplitude (here for demonstration we have considered $A_s=a/b$ and symbol frequency $f_s=0.15$ (as in Fig. 1a). For the FHN we used parameters: $a=0.8, \ b=0.7, \ \tau=10$. As a result the signal acquired the FHN-characteristic waveform, which we then attenuated to the parameters of the signal in Fig. 1a, i.e.: signal power $0.0065$ and mean $0.075$. Thus, we received a signal with the same power characteristics as in Fig. 1a but with neuromorphic modulation (NM), depicted in Fig. 1c. This signal was then inserted into the FHN processor (here no noise was added) - the output for NM signal has exhibited characteristic spiking even in the absence of noise, compare to the abscence of spiking for the signal with the same power characteristics in Fig. 1a. Moreover, the gain for NM signal (see black solid line  Fig. 1b) is comparable to that of the rectangular signal requiring 5 dB noise. The NM signal is less sensitive to the pre-modulation symbol frequency, although also peaks at $f_s=2f_0$. Thus, the FHN processor is modulation selective.

\section*{Results: Amplification via Noise Shaping}
The input shown in Fig. 1a is re-plotted in Fig. 2a: the signal (white) and noise when combined (blue) result in exciting the spiking regime (red), while in the spectral domain (see Fig. 2b) one can compare the clean input - signal (shown in black) and noisy input - signal plus noise (shown in blue) with the FHN output (shown in red): the inset highlights the output forming a peak in the vicinity of the FHN characteristic frequency $f_0$.

To quantify spike generation the spiking rate is used: a ratio of generated spikes to the number of encoded "ones" (see Fig. 2b). The spiking rate follows the exponential dependency on symbol frequency: in Fig. 2c see the spiking rate for various sets of  $\{$signal amplitude $A_s$; noise intensity $D$ $\}$    plotted in solid colored lines alongside the exponential approximation $exp(-A_1-A_2f_s)$ (where $A_1$ and $A_2$ are numerical constants) shown in dashed black lines. See in Fig. 2c the dependence of spiking rate on noise intensity for $f_0/f_s=1;2$ shown in blue and red lines. When the number of spikes is rescaled by $f_0/f_s$, the spiking rate converges  to the corresponding spiking rate for $f_0/f_s=1$ case. This is because the bandwidth of the FHN output saturates to the  FHN characteristic frequency value $f_0$ as shown in Fig. 2e (see  \cite{HFT}), i.e. there is a maximum number of spikes per symbol that can be supported.

To see into the underlying physical process let us consider a case of noise only excitation, i.e. when the FHN is fed only by noise. Compare the noise only case depicted in Fig. 2f to the excitation by noisy signal plotted previously in Fig. 2a, here all parameters are kept the same including the scales of the figures. The  noise at the output is drastically reduced (averaged out): so that compared to the input noise (see blue lines in Fig. 2a), the output (Fig. 2f) seems as a constant, also see the the enlarged inset. 

The output spectrum (plotted in Fig. 2g) reveals that the noise shaping effects takes place: the flat power spectrum of input noise is reshaped around the characteristic frequency and the tail follows $f^{-2}$-scaling law (see the supplemental note). When the input noise is of higher intensity (here 10 dB difference in magnitude is considered), the output spectrum is proportionally rescaled (compare blue and red lines in Fig. 2g, correspondingly).   
The effect occurs due to the FHN-induced noise filtering with memory, see the covariance for $<\eta(t),\eta(t+\Delta)>$ normalized by the noise variance for the output noise plotted in Fig. 2h). These two effects: (i) noise shaping (Fig. 2g) and (ii) memory  (Fig. 2h) result from the Harmonic frequency transformation \cite{HFT} and together enable selective signal amplification via constructive use of noise (see the supplemental note).


\section*{Conclusions}
We demonstrate that the FHN neuron may act as an amplifier by reshaping the input noise around the characteristic spiking frequencies enabling spiking even for below-the-threshold signals. Moreover, the FHN processor can selectively amplify: (i) signal distinguishing it from the noise and (ii) certain signal modulation waveforms. The effect is due to the harmonic frequency transformation, which enables the mechanism  of noise shaping (without the conventional requirement of the feedback loop).
 Our results pave the way for a paradigm change in the design of channel components for neuromorphic communication systems.

\begin{figure*}[!ht]
\centering
\includegraphics[width=\linewidth]{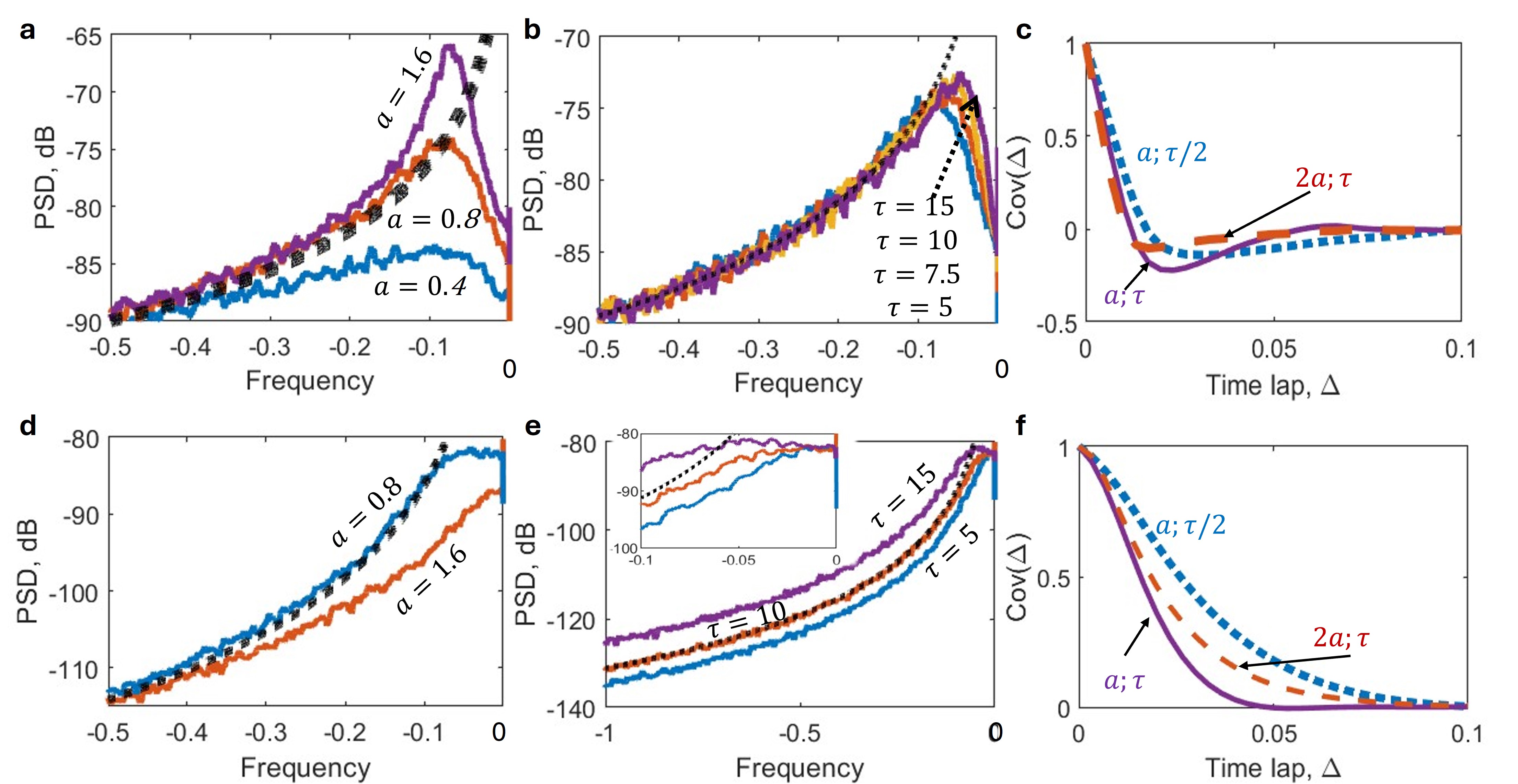}
\caption{\textbf{Noise shaping mechanism.}  
Two FHN variables act differently on the input noise: the fast variable (voltage, $v$), see PSD in \textbf{a-b}, exhibits noise localization and $f^{-2}$ tail (see the asymptote in dashed black lines), while the slow variable $w$ see PSD in \textbf{d-e}, exhibits flat-like dependency and $f^{-4}$ tail. Changing the FHN parameters enables to adapt the response accordingly: variation of $a/b$ enables to vary the noise squeezing strength (in panels \textbf{a,d} variables are plotted for fixed $\tau=5, b=0.8$ and varied $a$), while variation of $\tau$ allows to adjust the localization frequency:  in panels \textbf{b,e} variables are plotted for fixed $a=0.7, b=0.8$ and varied $\tau$). The difference between the fast and slow variables $v$ and $w$ can be seen via the normalized covariance  in panels \textbf{c,f}, correspondingly. Comparing the covariance for parameters $a=0.8; b=0.7; \tau=5$ and modified sets, illustrates the squeezing role of parameter $a/b$ and time-frequency scaling of $\tau$.}
\label{F2}
\end{figure*}

\section*{Supplemental Note }
\subsection*{Mechanism of noise shaping: two variables - two time scales}
Let us distinguish two main features of the observed noise shaping effect:

(i) the flat input noise is reshaped (accumulated) around the frequency that is within the FHN spiking range;

(ii) the tail of the output noise follows power law frequency scaling in PSD.

Pink noise is often used in noise shaping, for dithering and as a result of filtering, however noise accumulation at the spiking frequency represent a novel unusual phenomena. Usually, the system is designed so that the noise is moved to farther frequencies, separating it from the signal. 

In \cite{HFT} we have revealed that in the FHN model in the presence of within-the-threshold input the noise is accumulated at the spiking range frequencies (and higher order harmonics) creating a spectrum lobe inverse proportionate to the input signal bandwidth. This, in conjunction with the appropriate oversampling range, allows to achieve significant noise reduction and improvement of signal quality (see \cite{HFT}). 

For below-the-threshold signal (see the inset in Fig. 2b) we see that the output bandwidth is preserved to the characteristic spiking frequency, which is also the case for the noise only excitation (see Fig. 2g). Thus, the FHN process reshapes the noise accumulating it to the frequencies within the spiking range, thus assisting in transferring noise power to the signal.

The accumulated noise (caused by the interplay of two time scales from the slow and fast FHN variables) at the frequency of interest - spiking frequency plays a role of a pump  in parametric amplification process due to the cubic nonlinearity in the FHN (a three wave mixing process). While due to the  harmonic frequency transformation observed in the FHN \cite{HFT}, the process analogous to modulation instability occurs similar to the Faraday instability (FI) leading to the formation of spiking patterns.

Let us consider the noise shaping process in more detail: in Fig. 3 we consider excitation by noise only and compare the FHN variables: fast $v$ (panels a-c) and slow $w$ (panels d-f). It can be seen that the two variables exhibit drastically different behaviour in PSD: fast - has a clear peak at characteristic frequency and $f^{-2}$ tail, while slow - has a plateau like behaviour (see the inset in Fig. 3e), which is sharply cut at the characteristic frequency, after which $f^{-4}$ tail follows. Thus, the slow variable $w$ (see Fig. 3d,e)  performs averaging noise up to the characteristic frequency and squeezing noise  for higher frequencies - i.e. averaging noise as a frequency filter, while  the fast variable $v$ (see Fig. 3a,b) acts as a different filter, which localizes and amplifiers noise at a given frequency analogues a pump. Where by subsequent nonlinear interaction (similar to the three wave mixing), the signal is amplified and the spike is produced. 
Moreover, the covariance as a function of time lag further highlights the difference in the processes realised by these two variables. These two processes with different scales and dynamics allow to model complex temporal nonlinear interactions and feedback within a single RLC circuit and a simple equation.
Furthermore, changing parameters: $\tau$- for temporal scaling and $a/b$- for the noise shaping intensity enables to adapt the FHN processor to the signal and noise characteristic - changing the frequency of noise peak (via $\tau$) and relaxing requirement on noise intensity (via $a/b$). 

Thus, there is an interplay of two processes and two time scales, which leads to the noise shaping at the narrow range of the desired frequency - characteristic spiking frequency.

The tails with power-law scaling arise from the FHN acting as a matched filter, which results in a summation of noise components within  $f_0$ filter window. As the initial noise is a white  Gaussian noise, the result is the Brownian noise with $f^{-2}$ tail for $v$-variable and subsequent summation (filtering) of $v$-variable creates an $f^{-4}$-tail for the slow $w$-variable.

The peaks centered at the characteristic $f_0$-frequency arise from the Harmonic frequency transformation, which was first observed in the FHN model in \cite{HFT}. Let us consider here a simplified model, the output frequency is transformed as follows:
\[H(f)=f\Big[\frac{f_0}{f}\Big]\eta(f_0-f)+\frac{f}{\Big[\frac{f}{f_0}\Big]}\eta(f-f_0)\] (with $\eta$ being the Heaviside function). The HFT  results in frequency scaling for noise spectral components as:
\[\mathrm{HFT}(\eta_f) \rightarrow \eta_{f/n}|_{n=[f/f_0]}\]
Thus, the HFT results in additional summation process of noise components within the $f_0$-region as
\[\mathrm{HFT}(\eta_f) =\sum_{n=1}^\infty\int df \eta_{f/n}\delta_{n,[f/f_0]}e^{i f t/n}\]
Thus, the noise components at various frequencies are summed together, however the number of terms at each sum is different with the maximum for $f=f_0$ and decaying with the distance from it. This enables both formation of a peak at the vicinity of the characteristic frequency and correlation between the components.

%



\end{document}